\documentclass[%
 reprint,
 amsmath,amssymb,
 aps,
]{revtex4-1}

\usepackage{graphicx}% Include figure files
\usepackage{dcolumn}% Align table columns on decimal point
\usepackage{bm}% bold math
\usepackage{subfigure}
\usepackage{multirow}

\begin{document}

%\preprint{APS/123-QED}

\title{The TileCal Energy Reconstruction for LHC Run2 and Future Perspectives}%

\author{J. M. de Seixas, on behalf of the ATLAS Collaboration}
% \email{seixas@ufrj.br}
 %\altaffiliation[Also at ]{Federal University of Rio de Janeiro.}
%\author{L. M. de Andrade Filho}%
\affiliation{Federal University of Rio de Janeiro}
%\collaboration{ATLAS Collaboration}

%\author{L. M. de Andrade Filho}
% \email{luciano.andrade@engenharia.ufjf.br}
%\author{Bernardo S. Peralva}
%\email{bernardo@cern.ch}
%\affiliation{Federal University of Juiz de Fora}
%\affiliation{Third institution}
%\collaboration{on behalf of the ATLAS Collaboration}

\date{\today}% It is always \today, today,

\begin{abstract}
The TileCal is the main hadronic calorimeter of ATLAS and it covers the central part of the detector ($|\eta| \textless 1.6$). The energy deposited by the particles in TileCal is read out by approximately 10,000 channels. The signal provided by the readout electronics for each channel is digitized at 40~MHz and its amplitude is estimated by an optimal filtering algorithm. The increase of LHC luminosity leads to signal pile-up that deforms the signal of interest and compromises the amplitude estimation performance. This work presents the proposed algorithm for energy estimation during LHC Run 2. The method is based on the same approach used during LHC Run 1, namely the Optimal Filter. The only difference is that the signal baseline (pedestal) will be subtracted from the received digitized samples, while in Run 1 this quantity was estimated on an event-by-event basis. The pedestal value is estimated through special calibration runs and it is stored in a data base for online and offline usage. Additionally, the background covariance matrix will also be used for the computation of the Optimal Filter weights for high occupancy channels. The use of such information reduces the bias and uncertainties introduced by signal pile-up. The performance of the Optimal Filter version used in Run 1 and Run 2 is compared using Monte Carlo data. The efficiency achieved by the methods is shown in terms of error estimation, when different conditions of luminosity and occupancy are considered. Concerning future work, a new method based on linear signal deconvolution has been recently proposed and it is under validation. It could be used for Run 2 offline energy reconstruction and future upgrades.
\end{abstract}

\maketitle

%\tableofcontents

\section{\label{sec:introduction} Introduction}

\begin{figure}[h!]
	\centering
	\subfigure[]{
		\includegraphics[scale=.11]{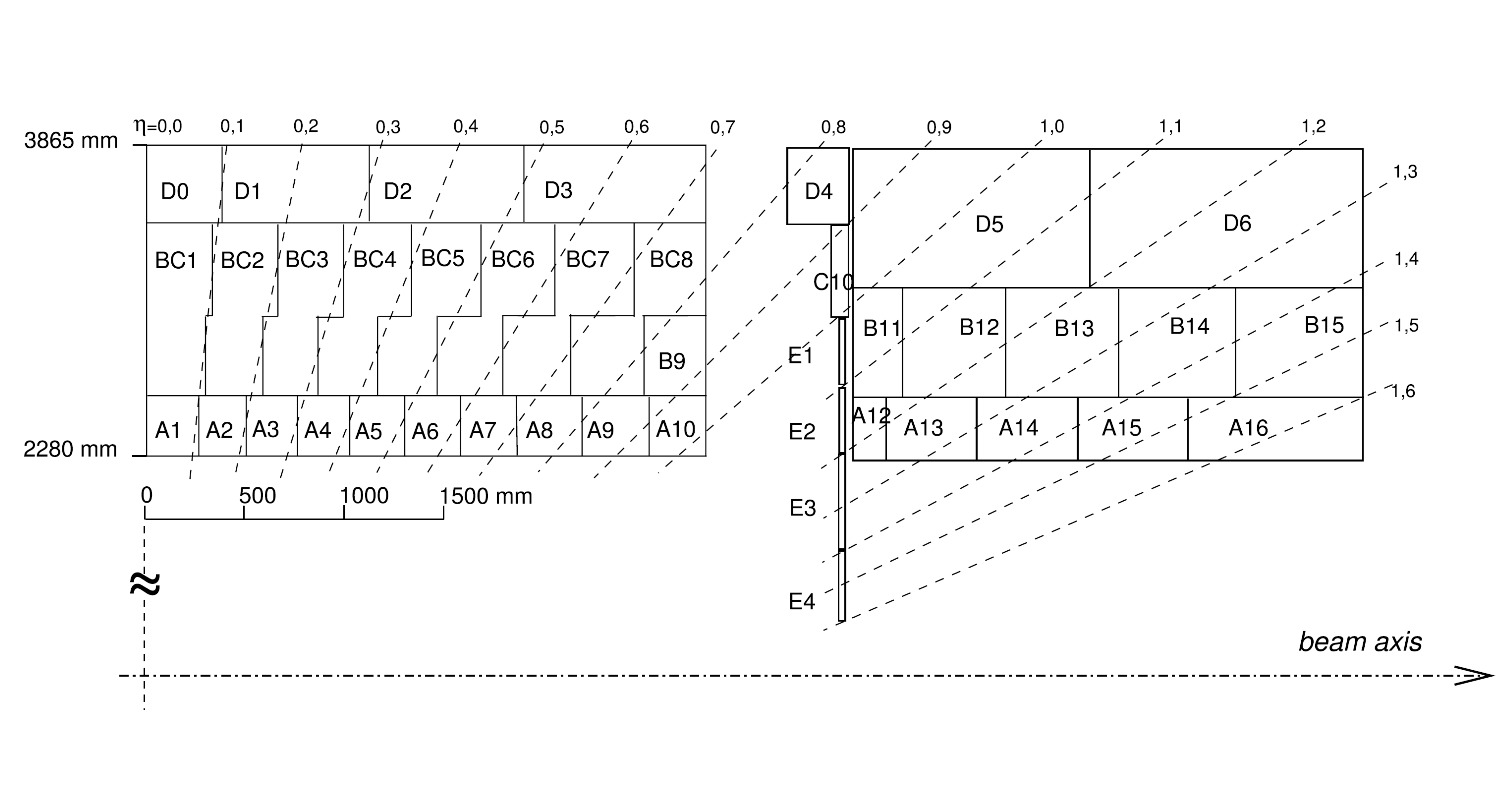}
		\label{fig:seg}
	}
	\subfigure[]{
		\includegraphics[scale=.3]{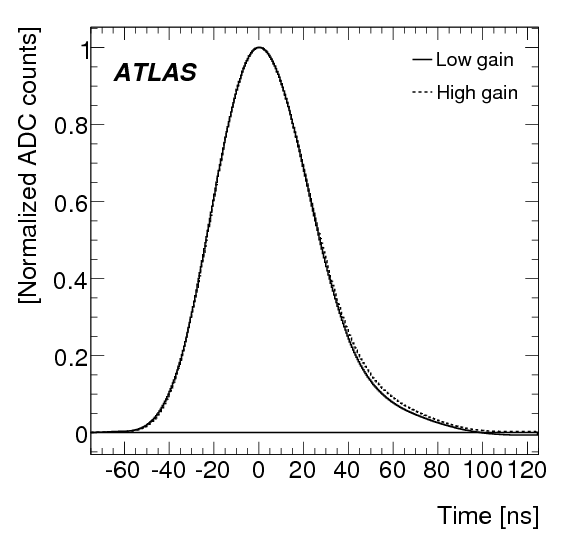}
		\label{fig:pulse}
	}
	\caption{(a) Schematic of the TileCal segmentation~\cite{atlas}; (b) Cell signal pulse shape after amplification and shaping~\cite{tile}.}
\end{figure}

The Tile Calorimeter (TileCal)~\cite{tile} is the central hadronic section of the ATLAS calorimeter and it covers the pseudo-rapidity ($\eta$) region laying at $|\eta| \textless 1.6$~\cite{atlas}. It consists of approximately 5000 cells each one readout by two channels. Radially, it is divided into 3 main layers, A, BC (B for the extended barrel) and D (see Figure~\ref{fig:seg}). Four special cells are located in the gap/crack region (E cells). The front-end electronics~\cite{3in1} shapes and amplifies the analog signals (two outputs with gain ratio of 1:64), providing a fixed and stable 150~ns pulse shape (see Figure~\ref{fig:pulse}). The pulse is sampled at 40~MHz and a readout window of 7 samples is available for digital processing for each channel. The signal arrival time and signal amplitude (energy) released in the channel are estimated online by a fast Optimal Filter~(OF) technique~\cite{of_orig}.

\begin{figure}[h!]
	\centering
	\includegraphics[scale=.35]{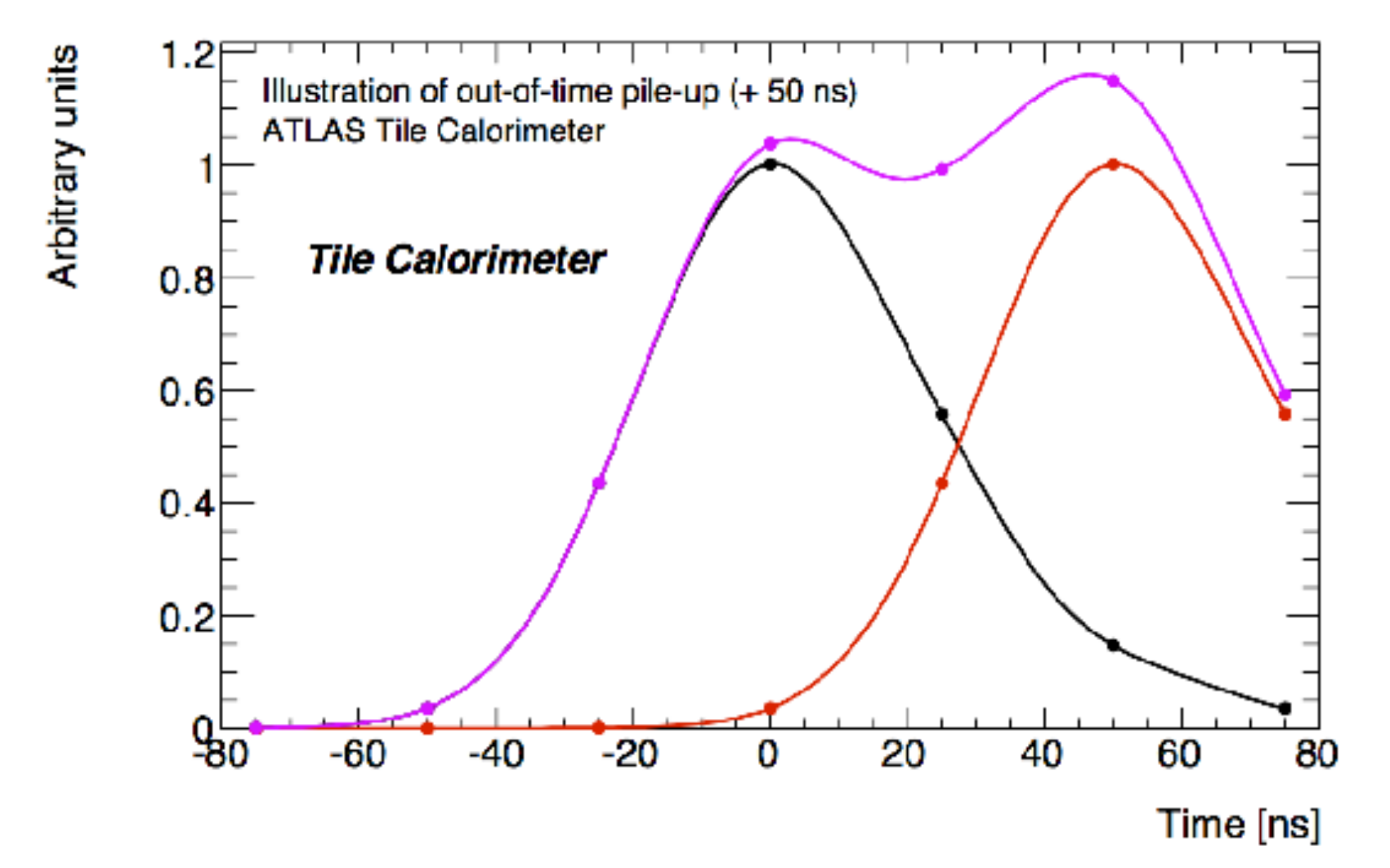}
	\caption{Illustration of the pile-up effect in TileCal~\cite{publicPlotsSimu}. The black curve corresponds to the signal of interest and the red curve is a out-of-time signal from a subsequent collision. The distorted signal is illustrated by the magenta curve.}
	\label{fig:pileup}
\end{figure}

The increase of the LHC luminosity will lead to higher occupancy in the ATLAS calorimeter system. Since the TileCal signal width and the readout window are larger than the LHC nominal event rate (25~ns), the cell signals may be generated by particles originated by interactions from subsequent proton-proton bunch-crossings. This process results in the observation of both in-time and out-of-time (OOT) signals within the same readout window. This effect is especially expected in cells from A and E layers of TileCal, which are more exposed and closer to the beam. Under pile-up conditions, the signal of interest located in the central bunch crossing is deformed (see Figure~\ref{fig:pileup}). As a result, the standard OF method becomes biased, since it assumes only the central bunch crossing signal corrupted with the usual electronic noise (WGN - White Gaussian Noise).

%The identity matrix showed to be a reasonable description of the noise under very low occupancy, where the background comprises mostly electronic noise. However, the OF technique foresees the use of the background covariance matrix in its design, which can handle at least the second order statistics of the background (pile-up plus electronic noise).

This work evaluates the use of the background covariance matrix in the design of the OF technique operating under pile-up conditions, and it introduces a discussion regarding future perspectives.

Next section introduces the OF method in details as well as an investigation of the computation of the background covariance matrix. Section~\ref{sec:results} shows the results using full ATLAS simulation Monte Carlo MinBias events. Section~\ref{sec:cof} introduces a new method to deal with the signal pile-up that has been studied. Finally, the conclusions are outlined in Section~\ref{sec:conclusions}.

\section{TileCal Optimal Filter method}
\label{sec:OF}

In TileCal, the digitized ADC samples $y_i$ can be expressed as~\cite{OFtile}:
\begin{equation}
	y_i = ped+Ag(t_i + \tau)+n_i
	\label{eq:signal}
\end{equation}
where $ped$ is the signal pedestal, $A$ is the true amplitude, $g(t)$ is the normalized reference pulse shape at time $t$ (output from shaper), $\tau$ corresponds to the phase between the expected and measured times, and $n_i$ is the background noise.

The online amplitude estimation needs to be simple and fast to meet Level-1 timing constraints~\cite{trigger}. The baseline method used in both hadronic~\cite{OFtile} and electromagnetic~\cite{OF} calorimeters in ATLAS is based on a weighted sum of the digitized samples, aiming at minimizing the variance in the estimation of the signal amplitude. In this approach, the estimate of the amplitude is given by:
\begin{equation}
	\hat{A} = \sum_{i=1}^{N} y_i w_i,
	\label{eq:of1}
\end{equation}
where $y_i$ are the digitized samples, $N$ is the number of samples available and the vector $w_i$ corresponds to the OF weights, which are computed offline.

The variance of the amplitude parameter to be minimized is given by:
\begin{equation}
	var(\hat{A}) = \textbf{w}^{T} \textbf{C} \textbf{w}
	\label{eq:of2}
\end{equation}
where $\textbf{w}=\{w_1, w_2, ..., w_N\}$ and \textbf{C} corresponds to the background covariance matrix.

The OF implementation that operated during Run1 in TileCal for both online and offline is called OF2, and it performs the optimization procedure subjected to the following constraints:
\begin{equation}
	\sum_{i=1}^{N} g_i w_i=1
	\label{eq:of3}
\end{equation}
\begin{equation}
	\sum_{i=1}^{N} g'_i w_i=0
	\label{eq:of4}
\end{equation}
\begin{equation}
	\sum_{i=1}^{N} w_i=0
	\label{eq:of5}
\end{equation}
where the vectors $g_i$ and $g'_i$ are the TileCal reference pulse shape and its derivative, respectively. The first constraint (Equation~\ref{eq:of3}) regards unbiased estimations, while the additional second and third constraints (Equations~\ref{eq:of4} and \ref{eq:of5}) are added to make the estimation procedure immune against phase and baseline fluctuations, respectively.

The weights $w_i$ can be found by solving the following matrix system:

\begin{equation}\label{of13}
	\left(
	{\begin{array}{cccccc}
			C_{1,1} & \dots & C_{1,7} & -g_{1} & -g_{2}' & -1     \\
			\vdots & \ddots& \vdots & \vdots & \vdots \\
			C_{7,1} & \dots & C_{7,7} & -g_{7} & -g_{7}' & -1     \\
			g_{1} & \dots & g_{7} & 0 & 0 & 0     \\
			g_{1}' & \dots & g_{7}' & 0 & 0 & 0     \\
			1 & \dots & 1 & 0 & 0 & 0     \\
		\end{array} } \right)
		\left(
		{\begin{array}{c}
				w_{1}        \\
				\vdots     \\
				w_{7}      \\
				\lambda    \\
				\xi        \\
				v          \\
			\end{array} } \right)=
			\left(
			{\begin{array}{c}
					0      \\
					\vdots \\
					0      \\
					1      \\
					0      \\
					0      \\
				\end{array} } \right)
			\end{equation}
where $\lambda,\xi,v$ are the Lagrange multipliers.
			
If the noise can be modeled as an uncorrelated Gaussian process, the covariance matrix \textbf{C} can be written as an identity matrix. This approximation holds if the pile-up noise is negligible.
			
If the third constraint (Equation~\ref{eq:of5}) is removed from the optimization procedure, we call this method OF1. The difference is that OF2 computes the baseline value on an event-by-event basis, while OF1 relies on the stability of the baseline, and it subtracts a fixed value from each ADC sample
\begin{equation}
				\hat{A} = \sum_{i=1}^{N} (y_i-ped) w_i.
				\label{eq:of6}
\end{equation}
The constant value $ped$ is computed through special runs and stored in a data base.
			
%It is worth mentioning that OF1 achieves similar performance than the Gaussian Matched Filter~(MF) based approach, which was recently proposed for TileCal energy estimation~\cite{MF}. The MF method is developed from the likelihood ratio test and, consequently, can be designed for any background model. However, for non-Gaussian background, its design leads to a nonlinear filter, which is impracticable to implement for LHC Run2.

%\section{Background description and OF weights computation}
%\label{sec:background}

%As described in the previous section, the OF method makes use of the expected pulse shape as well as the background covariance matrix (under the assumption to be modeled by a Gaussian distribution). This section presents a study on the description of the background under different luminosity conditions. We investigate the estimation of the background covariance matrix and the impact on the OF weights.

In the case where the effect of signal pile-up is not present, the background comprises mainly the electronic noise, which is often described by a White Gaussian process~\cite{peebles}. Under these conditions, the OF technique (both OF1 and OF2) operates on its optimal performance. However, the signal pile-up introduces another contribution to the background, which becomes non-Gaussian under such conditions. As a result, the OF becomes no longer optimal.

%Channel~10 of EBA~43 (cell A13) was used to illustrate this study as it presents higher occupancy. The cell occupancy scales with the $<\mu>$ value, and in this particular cell we expect an actual occupancy of approximately $10\%$ for $<\mu>=40$~\cite{noteJimmy}.

%\begin{figure}[ht]
%	\centering
%	\subfigure[]{
%		\includegraphics[scale=.36]{GaussianMu0}
%		\label{fig:noiseMu0}
%	}
%	\subfigure[]{
%		\includegraphics[scale=.36]{GaussianMu40}
%		\label{fig:noiseMu40}
%	}
%	\subfigure[]{
%		\includegraphics[scale=.36]{FitMu0inMu40}
%		\label{fig:noiseMu0inMu40}
%	}
%	\caption{Distribution of the incoming ADC samples for $<\mu>=0$ (a), $<\mu>=40$ (b) and the Fit for $<\mu>=0$ superimposed on the histogram of $<\mu>=40$ (c) for channel 10 (cell A13).}
%\end{figure}

The correct treatment of the pile-up as background would require a non-linear modeling, which is impracticable for LHC Run2 due to hardware limitations. However, the background covariance matrix can be used in the OF design in order to reduce the uncertainties and bias due to the pile-up, improving the energy estimation performance of the OF method.

The occupancy of most of the TileCal cells is low, and the pile-up signal can be treated as outlier for the majority of the cells. Additionally, since the covariance matrix is very sensitive to outliers, alternative ways of computing this quantity must be considered.

The classical covariance matrix estimation~\cite{peebles} takes into account the whole dataset as follows:
\begin{equation}
	cov(x,k) = \frac{\sum_{i,j=1}^{M}(x_i-\bar{x}) \times (k_j-\bar{k})}{M},
	\label{eq:covariance}
\end{equation}
where $\bar{x}$ and $\bar{k}$ are the mean values of the random variables $x$ and $k$, respectively, evaluated at instant $i$ and $j$. $M$ is the total number of events in the given dataset. The outliers result in a larger variance which in turn lead to biases in the computation of the OF weights.

Alternatively, the Minimum Covariance Determinant Estimator~(MCDE)~\cite{mcde} algorithm randomly takes a subset of the background events and computes its classical covariance matrix (through Equation~\ref{eq:covariance}) and its determinant. The algorithm repeats this procedure 500 times (default) and it selects the subset that resulted in the lowest determinant. This subset contains the events that has the lowest covariance between the samples (ADC digits) and therefore it consists of the most probable events, disgarding most of the outliers (high energy pile-up in this case).

\section {Simulation Results}
\label{sec:results}
The performance evaluation will be carried out using full ATLAS Minimum Bias~(MB) simulation~\cite{simu}. Both OF1 and OF2 algorithms are applied to this data sample to estimate the signal reconstruction performance. Since the MB events contain only background (electronic plus pile-up), the mean and the dispersion of the energy distribution can be used to quantify the bias and the estimation error, respectively. Also, since the OF method is a linear approach, this estimation error and bias are the same for any energy range.

\begin{figure}[h!]
	\centering
    \includegraphics[scale=.25]{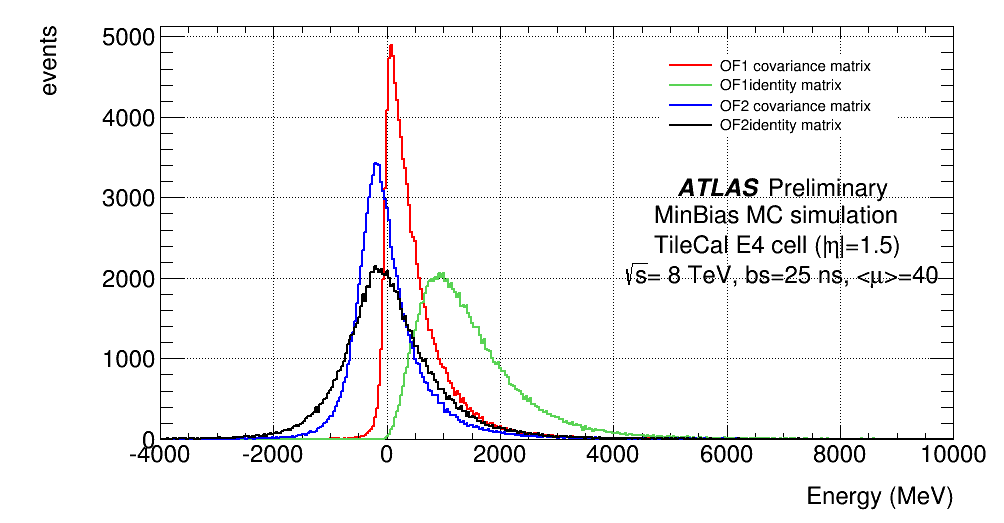}
	\caption{Energy distributions for a high occupancy cell (E4) using different covariance matrices for OF1 and OF2 methods~\cite{publicPlotsSimu}.}
	\label{fig:histE4final}
\end{figure}

Figure~\ref{fig:histE4final} shows the reconstructed energy from the MB events for a high occupancy cell, namely the E4. It can be noted that the use of the covariance matrix improves considerably the energy resolution. The OF1 with covariance matrix shows a sharp negative and positive tails, whereas OF2 presents a large dispersion in both negative and positive tails. It can also be seen that the use of the identity matrix in the OF1 design leads to a large positive bias for such high occupancy cells in TileCal.

%\begin{figure}[h!]
%	\centering
%	\subfigure[]{
%		\includegraphics[scale=.44]{E3histFinal}
%		\label{fig:histE3}
%	}
%	\subfigure[]{
%		\includegraphics[scale=.44]{E4histFinal}
%		\label{fig:histE4final}
%	}
%	\caption{Energy distributions for high occupancy cells using different covariance matrices for OF1 and OF2 methods.}
%\end{figure}

\begin{figure}[h!]
	\centering
	\includegraphics[scale=.26]{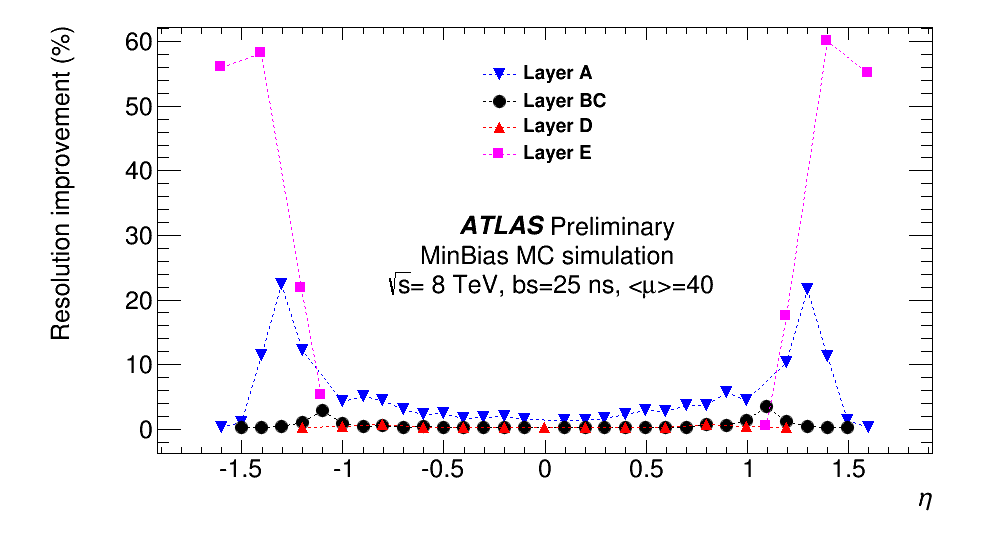}
	\caption{Resolution Improvement (RI) for OF1 as a function of pseudorapidity for the four TileCal layers~\cite{publicPlotsSimu}. The E4 cell is located at Layer E and $|\eta| = 1.6$.}
	\label{fig:OF1vsOF1}
\end{figure}

The improvement, in percentage, in the estimation error (RMS of the distribution) that is introduced in the OF1 estimation by using the correct covariance matrix (computed through MCDE) with respect to the identity matrix is quantified by
\begin{equation}
	RI (\%) = 100 - \frac{RMS_{covariance}}{RMS_{identity}} \times 100,
	\label{eq:improvement}
\end{equation}
where $RI$ stands for Resolution Improvement.
Figure~\ref{fig:OF1vsOF1} shows the RI in all cells in the central and extended barrels for $\phi=4.2$~rad. As expected, the A and E cells in the extended barrels present the largest improvements due to their higher occupancy.

The use of the pedestal value, retrieved form a data base, does not introduce bias to the final estimate for all cells as the mean of the distributions are smaller than 1~ADC count (1~ADC counts corresponds to approximately 12~MeV for a high gain channel). The cells located in the BC and D layers presented similar mean and RMS values, for both OF1 with covariance and identity matrices due to their low occupancy. The OF2 shows improvements by using the covariance matrix with respect to identity matrix only for the E3 and E4 cells.

\section {Future perspectives}
\label{sec:cof}
The OF1 method incorporates the signal pile-up through the use of the background covariance matrix. Although the background becomes non-gaussian under pile-up conditions and the OF design remains luminosity-dependent, such approach still presents good performance and implementation simplicity. For the case of high pile-up conditions (cell occupancy greater than $10$\%, for instance), an alternative technique has been developed. The Constrained Optimal Filter (COF) is a method based on signal deconvolution that recovers the energies from all signals within the readout window.

\begin{figure}[h!]
	\centering
	\includegraphics[scale=.4]{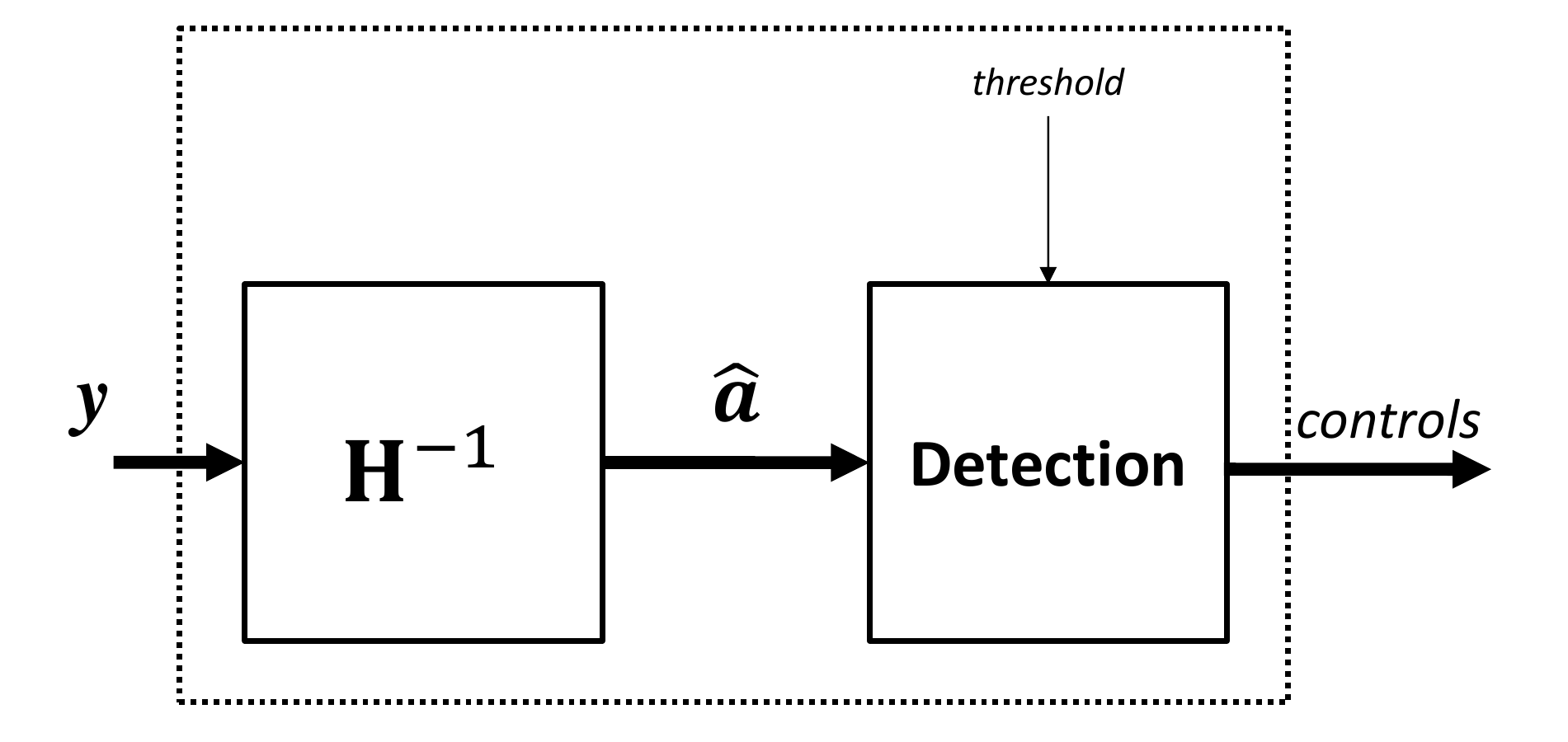}
	\caption{Pile-up detection for the COF method.}
	\label{fig:pileupdet}
\end{figure}

The COF is a two-step algorithm. Firstly, it detects the presence of OOT signals by estimating the amplitude of the signals within the readout window. The estimation is performed by the following equation:
\begin{equation}
	\hat{\textbf{a}}=\textbf{H}^{-1}\textbf{y}
	\label{eq:dm}
\end{equation}
where $\textbf{H}$ is a $7 \times 7$ matrix where each row corresponds to a shifted version of the reference pulse shape. It should be stressed that the amplitude estimation is independent from the luminosity. The decision is based on a simple threshold, which is defined according to the cell noise energy (see Figure~\ref{fig:pileupdet}).
Secondly, it designs an optimal filter where the constraints are set to estimate the amplitude of only the detected signals.

\begin{figure}[h!]
	\centering
	\includegraphics[scale=.37]{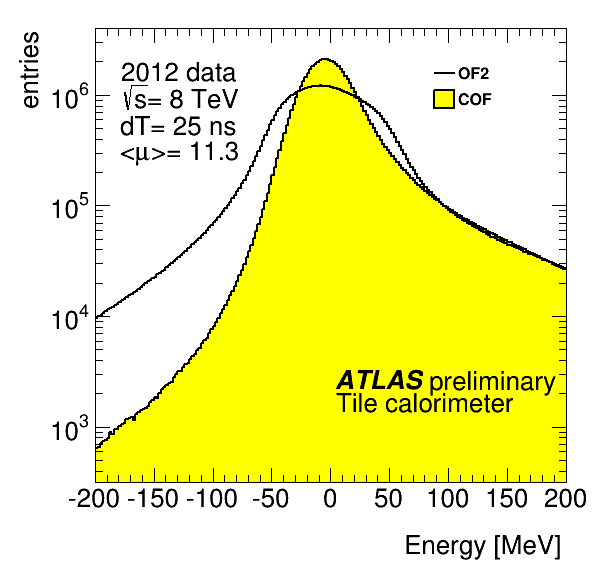}
	\caption{Cell energy distribution reconstructed by OF2 and COF around noise region ($\pm200$~ns)~\cite{publicPlotsReal}.}
	\label{fig:eneHist}
\end{figure}

In order to show the performance of the COF method, Figure~\ref{fig:eneHist} shows the distributions of the cell energy reconstructed by the OF2 and COF methods using p-p collision data with 25~ns of bunch spacing (dT) and maximum average number of interactions per crossing ($<\mu>$) of 11.3. Under these conditions, the energy reconstructed by COF is resilient to OOT signals. It can be seen that OF2 presents a large negative tail (bias) due to the presence of OOT signals located at the $\pm$75 and $\pm$50~ns. Since for COF the OOT signals within the readout window do not spoil the estimation of the signal of interest, it presents better performance (smaller dispersion) under pile-up conditions than OF2.

\section {Conclusions}
\label{sec:conclusions}

We presented a study on using a covariance matrix computation for absorbing the pileup noise and thus improving energy reconstruction in the TileCal. The results show that the OF1 method taking the pedestal value from the data base and using such covariance matrix is the most suitable method for online energy reconstruction in TileCal for LHC Run2. The pedestal value is computed through special pedestal runs, and both the covariance matrix and the OF weights are computed offline, and loaded into the digital signal processor available in the Readout Driver for online energy estimation.

For high-luminostiy conditions, where the signal pile-up is present, an alternative method (COF) based on signal deconvolution is being tested within TileCal, showing promising results.

\begin{acknowledgments}

The financial support form CAPES, CNPq, FA\~{A}ˆRJ, FAPEMIG, RENAFAE (Brazil) and the European Union (E-Planet project) is acknowledged.

\end{acknowledgments}

\nocite{*}


\begin{thebibliography}{90}

%\addcontentsline{toc}{chapter}{Refer{{\accent 94 e}}ncias Bibliogr\'aficas}

%\centerline{\bf  REFERENCES}

%\vspace*{5mm}

\bibitem{tile}
ATLAS Collaboration, \emph{Readiness of the ATLAS Tile Calorimeter for LHC collisions}, EPJC, \textbf{70}, pp 1193-1236, 2010.

\bibitem{atlas}
ATLAS Collaboration, \emph{The ATLAS Experiment at the CERN Large Hadron Collider}, JINST, \textbf{3}, S08003, 2008.

\bibitem{3in1}
K. Anderson, \emph{et al}, \emph{Front-end Electronics for the ATLAS Tile Calorimeter}, Proceedings of Fourth Workshop on Electronics for LHC Experiments, 1998.

\bibitem{of_orig}
G. Bertuccio, E. Gatti and M. Sapietro, \emph{Sampling and optimum data processing of detector signals}, Nuclear Instruments and Methods in Physics Research A, v. 322, 271 - 279, 1992.

%\bibitem{pileup}
%Clement, C. and Klimek, P., \emph{Identification of pile-up using the quality factor of pulse shapes in the ATLAS Tile Calorimeter}, IEEE Nuclear Science Symposium and Medical Imaging Conference (NSS/MIC), pp 1188-1193, 2011.

\bibitem{publicPlotsSimu}
ATLAS Collaboration, \url{https://twiki.cern.ch/twiki/bin/view/AtlasPublic/ApprovedPlotsTileReconstruction}.

\bibitem{OFtile}
E. Fullana, \emph{et al}, \emph{Digital Signal Reconstruction in the ATLAS Hadronic Tile Calorimeter}, IEEE Transactions On Nuclear Science, \textbf{53} 4, 2006.

\bibitem{trigger}
Lundberg, J. et al., \emph{Performance of the ATLAS first-level trigger with first LHC data}, Real Time Conference (RT), 17th IEEE-NPSS, 2010.

\bibitem{OF}
Cleland, W. E. and Stern, E. G., \emph{Signal processing considerations for liquid ionization calorimeters in a high rate environment}, Nuclear Instruments and Methods in Physics Research, A338 467-497, 1994.

%\bibitem{MF}
%B. Peralva, \emph{et al}, \emph{The TileCal Energy Estimation for Collision Data Using the Matched Filter} IEEE Nuclear Science Symposium and Medical Imaging Conference pp 1-6, 2013.

\bibitem{peebles}
P. Peebles, \emph{Probability, Random Variables, and Random Signal Principles}, McGraw-Hill, 2000.

\bibitem{mcde}
P. Rousseeuw and K. Dreissen. \emph{A Fast Algorithm for the Minimum covariance Determinant Estimator}, Technometrics, Vol. 41, No. 3, 1999.

\bibitem{simu}
J. Chapman, \emph{ATLAS Simulation Computing Performance and Pile-Up Simulation in ATLAS} LPCC Detector Simulation Workshop, 2011.

%\bibitem{noteJimmy}
%G. Lima and J. Proudfoot, \emph{Studies of cell response using occupancy rates measured in Tile 2 Calorimeter ZeroBias data with 50ns bunch-spacing}, ATL-COM-TILECAL-2013-032, 2013.

\bibitem{publicPlotsReal}
ATLAS Collaboration, \url{https://twiki.cern.ch/twiki/bin/view/AtlasPublic/TileCaloPublicResults}.

\end{thebibliography}

%\input{bibliography}

\end{document}